\documentclass[14pt,leqno]{extarticle}

\usepackage{graphicx}
\usepackage{indentfirst,csquotes}
\usepackage{spverbatim}
\usepackage{booktabs}
\usepackage{amssymb,amsthm,amsmath}
\usepackage{xcolor,paralist,hyperref,fancyhdr,etoolbox}
\usepackage{rotating}
\begin{document}
\title{\textbf{Applying an Agentic Coding Tool for Improving Published Algorithm Implementations}} 
\author{Worasait Suwannik}
\date{\today}
\maketitle

\begin{abstract}
We present a two-stage pipeline for AI-assisted improvement of published algorithm implementations. In the first stage, a large language model with research capabilities identifies recently published algorithms satisfying explicit experimental criteria. In the second stage, Claude Code is given a prompt to reproduce the reported baseline and then iterate an improvement process. We apply this pipeline to published algorithm implementations spanning multiple research domains. Claude Code reported that all eleven experiments yielded improvements.  Each improvement could be achieved within a single working day. We analyse the human contributions that remain indispensable, including selecting the target, verifying experimental validity, assessing novelty and impact, providing computational resources, and writing with appropriate AI-use disclosure.  Finally, we discuss implications for peer review and academic publishing.
\end{abstract} 



\section{Introduction}

Agentic coding refers to a paradigm in software development in which autonomous or semi-autonomous software agents, often powered by large language models, are capable of planning, generating, executing, and iteratively refining code with minimal human intervention. These systems can also interact with external command line tools and manage dependencies by installing library packages, thereby extending their operational scope beyond static code generation.

The growing practice of releasing open-source implementations and publicly available datasets alongside research papers creates a significant opportunity for these capabilities. Each such release provides a self-contained experimental environment: the code is runnable, the data is accessible, and the baseline performance is documented. These are precisely the conditions under which agentic coding tools can operate effectively, yet their application to improving published algorithm implementations remains largely unexplored.

This paper demonstrates that agentic coding tools can improve published algorithm implementations.  Our central contribution is a two-stage pipeline for AI-assisted improvement of published algorithm implementations. In the discovery stage, a commercial LLM with deep-search capabilities is prompted to identify recent, well-scoped papers satisfying explicit experimental criteria covering recency, journal tier, code availability, and execution time constraint. In the improvement stage, Claude Code is given a structured prompt instructing it to reproduce the paper's baseline and then iteratively improve upon it, saving each attempt as a numbered script and documentation file, for a maximum of twenty runs. 

We apply this pipeline to eleven published algorithm implementations spanning eleven distinct research domains: combinatorial optimization, explainable AI, pattern mining, image segmentation, data streaming, distributed systems, network security, graph machine learning, molecular simulation, computational physics, and bioinformatics. The pipeline achieved improvements in all eleven cases. Beyond the experimental results, we examine what the pipeline reveals about the evolving role of human researchers: not as coders, but as domain selectors, novelty evaluators, impact assessors, resource providers, experimental verifiers, and communicators.

The remainder of this paper is organized as follows. Section 2 surveys related work. Section 3 describes our methodology. Section 4 presents results across all eleven domains. Section 5 discusses strength of AI, limitations from our prompt design and the nature of agentic assistants, the human role, and implications for peer review and publishing. Section 6 concludes.

\section{Related Work}
AI has been applied across multiple stages of the scientific research process, including literature retrieval, hypothesis generation, experiment design, data analysis, manuscript drafting, and peer review. One line of work attempts to operationalize this vision end-to-end; ResearchAgent \cite{researchagent}, for instance, automates problem formulation, experimental design, and structured review.  Its output was evaluated against human assessors on dimensions such as problem significance, methodological validity, and experiment reproducibility. 

Several works demonstrate that coding agents can drive research progress through iterative refinement. Claudini \cite{claudini} shows that supplying existing adversarial attack implementations to an agent that autonomously iterates on them, without human intervention, yields a method that outperforms more than 30 prior approaches. AutoResearch \cite{autoresearch} similarly structures the research process as a continuous loop, using a minimal framework of a single bootstrap file and one writable working file, driven by a prompt that instructs the agent to repeatedly run, improve, and log results, demonstrating that a lightweight agentic framework is sufficient to enable effective experimentation.

Coding agents have also been applied to code transformation and optimization tasks. GeoFM \cite{geofem} translates CPU-parallelized Fortran code into CUDA GPU code, treating existing directives as structural hints. PostTrainBench \cite{posttrainbench} incorporates Claude Code into post-training pipelines for turning base LLMs into instruction-following assistants.

Agentic coding has been applied to open mathematical research. In \cite{claudecycle}, a researcher applies a coding agent to an open problem by having it construct candidate structures and test them computationally. Since exhaustive verification is infeasible, the agent's role is not to produce a proof but to surface patterns that human researchers can build upon. This work inspired the iterative refinement approach used in our improvement prompt.

\section{Method}
Our pipeline comprises two stages. Both stages rely on prompts that were refined through experience across the eleven experiments; we describe the final prompts and the reasoning behind each design choice.

\subsection{Discovery Stage}

The objective of the discovery stage is to identify a published algorithm or method suitable for AI-assisted improvement. We use ChatGPT's Deep Research, an AI research tool capable of browsing the web and cross-referencing sources to produce a report. Such tools can evaluate hundreds of online sources and synthesize the findings within 5 to 30 minutes \cite{openai2025deepresearch}, including cited reports, a task that would take a human researcher hours of manual search. We issue the following prompt using ChatGPT's Deep Research feature, substituting the target domain as needed:

\begin{minipage}{\textwidth}
\begin{spverbatim}
Locate a [DOMAIN] research paper that satisfies all of the following  criteria:
  - The paper is accompanied by Python or C++ implementation available on GitHub. 
  - The associated dataset(s) are publicly available for download and can be used for testing.
  - The provided code can execute on at least two datasets within 30 seconds.
  - Published after 2021 in a Q1 or Q2 journal.
  
\end{spverbatim}
\end{minipage}

The rationale for each criterion is as follows:

\begin{itemize}
    \item \textbf{GitHub implementation} Ensures reproducibility by eliminating papers where the reported algorithm cannot be independently verified or where subtle implementation differences make replication ambiguous.
    
    \item \textbf{Publicly available dataset.} Ensures that any improvement we report can be evaluated by third parties without data access requests.
    
    \item \textbf{30-second execution bound.} With up to twenty iterative improvement attempts per experiment, a single run exceeding thirty seconds would make the full exploration slow. This criterion was adjusted in the machine learning experiment where model training substantially exceeds this threshold.
    
    \item \textbf{Q1 to Q2 journal, post-2021.} Ensures the paper has passed rigorous peer review while reducing the likelihood that improvement strategies for the specific work have already appeared in the model's training data.
\end{itemize}

\textbf{Note.} The combinatorial optimization experiment (the first in Table~\ref{tab:results}) did not use this discovery stage. The author became aware of that paper when students presented the work in a seminar class. The discovery stage was adopted for all subsequent experiments.

\subsection{Improvement Stage}

Once a target paper was identified, we  downloaded the PDF, renamed it baseline.pdf, and placed it in the Claude Code working folder. We then opened a Claude Code session (starting with Claude Sonnet 4.6, with the researcher switching to Claude Opus 4.6 when usage limits were reached on several occasions, and Claude Code 2.1.96) and issued the following prompt:

\begin{minipage}{\textwidth}
\begin{spverbatim}


Read the paper baseline.pdf.  Select a metric-dataset pair that allows for rapid experimentation and provides clear potential for improvement.  Reproduce the paper's baseline for this selection using the provided source code or dataset, if available.  Develop a more effective implementation and iteratively refine it for up to 20 runs.  In each run, save the implementation as exploreXX.py, save experimental results as resultXX.csv, and document the process in planXX.md.  Each planXX.md must include the hypothesis, core idea, changes made from the previous run, comparison with prior results and the paper's baseline, and next steps.  Stop immediately once a result surpasses the paper's reported performance on the selected metric and dataset.  Summarize key interactions and key findings in summary.md.

\end{spverbatim}
\end{minipage}

The rationale for each element of the improvement prompt is as follows:

\begin{itemize}
    
    \item \textbf{Select a metric-dataset pair.} Delegates the  decision to the model: it reads the paper, identifies which table or figure is most promising, and proposes a focused improvement target.
    
    \item \textbf{Reproduce the paper's baseline.} Ensures the model runs the existing code before proposing any modifications, and surfaces discrepancies between the paper's reported results and the code's actual output.
    
    \item \textbf{Up to 20 runs.} Prevents the model from running indefinitely when the improvement target is not reachable.  
    
    \item \textbf{Auditable research log.} Creates an auditable research log (exploreXX.py, planXX.md, resultXX.csv, summary.md) that allows retrospective analysis of which hypotheses were tried and why. Key interactions are recorded to reveal how the human guided or intervened.  Code is saved with the same file extension as the baseline code to ensure a fair comparison. 
    
    \item \textbf{Immediate stop condition.} Terminates exploration as soon as the improvement criterion is met, avoiding unnecessary computation.
\end{itemize}

We wrote this prompt in one paragraph, inspired in part by the iterative refinement prompt used in \cite{claudecycle}. Compare to autoresearch's program.md, a baseline instructions for an agent, that has sections, numbered items, output format and specific to ML model development \cite{autoresearch}.

\subsection{Prompt Evolution}
The prompts above represent the final versions after several refinements drawn from observations across early experiments. Notable changes included adding the resultXX.csv requirement so the researcher could review  results when directing subsequent runs, adding summary.md for a consolidated record of key findings and interactions, and restricting implementations to Python and C++ to avoid requiring additional compiler installations.

\section{Results}

Table \ref{tab:results} summarizes the eleven experiments. In every case, Claude Code reported that its method surpassed the paper's reported baseline on the selected metric, and every AI reported improvement was achieved within a single working day. Across experiments it tested different approaches and documented why each succeeded or fell short. The prompt required each run to be documented with an explanation of what had
worked, what had not, and why. Each subsequent run showed awareness of prior results, producing an iterative loop that resembles the structure of a small research project. 

In the pattern mining experiment, the AI compared against the classical baseline rather than the paper's proposed method, because on the selected dataset the classical method substantially outperforms the proposed one in both pattern count and runtime. The discrepancy in pattern count between the GitHub code and the paper's reported values suggests the repository may not correspond exactly to the version evaluated in the paper. The AI's improved implementation ran 6.4 times faster.

In the network security experiment, the defense success rate more than doubled in a calibrated simulation approximating the paper's setup. However, two limitations temper this claim. First, the defense success rate is drawn from 50 benchmark networks against the paper's 200. Second, the paper's simulation engine is proprietary, so the comparison is between our simplified Python surrogate and the paper's original results. Without access to the original engine, a fully rigorous comparison is not possible.

In the graph ML experiment, improvement was achieved by replacing the paper's method with an alternative approach rather than modifying it. Moreover, this experiment went to 38 iterations because at 20th iteration, it cannot outperform the baseline.  Even though the prompt mention up to 20 run and AI wrote in plan20.md with title ``Final Comprehensive Approach"  but AI's own prioritization of performance goal over iteration cap cause AI to keep experimenting. 

The computational physics paper describes a library with a performance bottleneck in its primary conversion function. The bottleneck arose because the code solved a harder problem than necessary. The fix avoids the expensive intermediate computation entirely. According to an unverified AI-generated report, this reduces runtime on realistic inputs, with outputs matching the original on all tested cases. Only one standard test case shows a minor regression.

\begin{sidewaystable}
\centering
\caption{Selected Single-Metric, Single-Dataset Result Across Eleven Experiments.  * items are explained in the text}
\label{tab:results}
\small
\begin{tabular}{llll}
\toprule
Domain & Metric & Improvement & Source \\
\midrule
Combinatorial Opt. & Runtime & 193× faster; consistent near-optimal & New algorithm \\
Explainable AI & Sparsity  & Lower sparsity & New algorithm  \\
Pattern Mining* & Runtime & 6.4x faster & Data representation \\
Image Segmentation  & Quality at high $K$& Global optimum; $>1000\times$ faster    & New algorithm \\
Data Streaming & F1 at 10 KB & Higher F1 & Compact Representation \\
Distributed Systems & Lookup latency     & $>2\times$ faster & Data structure \\
Network Security* & Defense success rate & More than doubled & New algorithm \\
Graph ML* & Accuracy & 8\% & Algorithm replacement \\
Molecular Sim.      & Kinetic temp. error& Error reduced at all tested values & Momentum Estimator Improvement \\
Computational Phys* & Runtime   & Runtime reduced                        & Code optimization \\
Bioinformatics      & Runtime & 10.5$\times$ typical; 34.3$\times$ large   & algorithmic redesign \\
\bottomrule
\end{tabular}
\end{sidewaystable}

\section{Discussion}

\subsection{Limitations: What This Study Did Not Do}

Transparency about the scope of our experimental validation is essential. The following limitations apply to all eleven experiments.

\textbf{Independent code verification.} We did not conduct a line-by-line audit of Claude Code's implementations. We accepted the model's reported results without running the code outside Claude Code REPL (read–eval–print loop) shell. The results should therefore be treated as preliminary and requiring replication and careful examination before formal publication.

\textbf{Reproducing the experiment.} Reproducing a published experiment is more difficult than it appears, even when code and data are publicly available. In our experiments, these challenges included a proprietary simulation engine that required building a surrogate from scratch, author code that diverged from the paper's written description, and initial reimplementations that produced results inconsistent with reported values. We found that AI can substantially aid this process: implementing algorithms from paper descriptions, cross-checking results against reported tables to detect formula errors, inspecting author code to resolve paper–implementation discrepancies, and iteratively calibrating a surrogate when the original engine is unavailable. This suggests AI-assisted reproduction could become an important step in the replication pipeline, reducing the burden on human researchers who validate published results.

\textbf{Metric and dataset coverage.} Claude Code improved the selected metric but did not necessarily improve all metrics reported in the paper. In some experiments this created tradeoffs: a method that gained on one measure lost ground on another. Dataset scope also evolved across runs. In some cases the human researcher judged the small dataset in early run too unreliable as a benchmark and directed a switch to the another dataset before continuing.

\textbf{Novelty verification.} We did not  verify whether each proposed improvement has been independently published elsewhere. This possibility is non-trivial.  The baseline papers were published after 2021, and between their publication and the present (2026) independent improvements may exist that Claude Code would not recognize as prior work. A complete academic submission therefore requires a targeted literature review for each proposed technique, conducted by a human researcher with domain knowledge.

\subsection{Strengths of Claude Code}

Claude Code demonstrated a good command of programming languages, profiling bottlenecks, code optimization, fixing runtime exceptions, prowess in command-line tools, working across machine learning frameworks, and selecting among algorithmic alternatives. AI selected reasonable metric-dataset pairs from the baseline paper without guidance, and produced well-structured hypothesis-driven documentation for each run. It did not simply make the nearest available change but instead explored qualitatively different approaches across runs, switching strategies when a direction reached diminishing returns. In one experiment it could identified structural properties of the problem, such as separability or storage inefficiency, that the original authors had not exploited. These insights were not prompted by the human researcher; they emerged from the model's own analysis.

\subsection{Prompt Limitations and the Nature of Agentic Assistance}\label{section_prompt_limitation}

\textbf{Prompt Limitations.} The prompt as written focused solely on improving performance and said nothing about the integrity checks that a careful researcher would apply routinely in each domain. Several observed weaknesses trace directly to this omission. The instruction to stop immediately once a result surpasses the baseline caused the model to halt at the first improvement, even when that improvement was marginal. The broad framing of ``select a metric-dataset pair'' gave the model full latitude over dataset choice, leading in some experiments to early runs on a small dataset followed by a human-initiated switch to another dataset. Finally, the instruction to ``outperform the reported results'' without specifying the exact target row created ambiguity in at least one experiment where the paper's own proposed algorithm was outperformed by another baseline  reported in the same paper, and the misidentification only became apparent later. These prompt-level issues are, in principle, correctable with more precise instructions.  The harder problem is with failures the prompt author did not anticipate: the space of possible errors in any given domain is rarely fully enumerable in advance, and a model cannot flag what it does not know it is missing.

\textbf{The nature of agentic assistance.} These observations reflect a fundamental tradeoff in how agentic assistants are designed. A brief, general prompt achieves broad usability.  The model handles implementation, tool use, and documentation autonomously with minimal human direction. The cost is precision.  An agentic assistant trades some of that precision for speed and accessibility. 

\subsection{The Human Role in AI-Assisted Research}
\label{section_human_role}
The eleven experiments collectively illustrate what human researchers must contribute when Claude Code does the implementation. AI coding assistants possess capabilities that overlap substantially with those traditionally expected of research assistants and junior researchers in the field of computer science and related computational fields: rapid implementation of algorithms, fluency with command-line tools, knowledge of machine learning frameworks, familiarity with data structures and algorithm design, code profiling, various language and platform. As these capabilities mature, the skills most valued in a research assistant may shift away from implementation and toward the higher-order contributions described below.

\textbf{Critical thinking and experimental verification.} The most important and recurring human contribution was applying critical judgment to the model's outputs. As the cases  misidentified dataset files and used a synthetic dataset, AI produced plausible-looking results without flagging that anything might need checking. The burden of skepticism fell consistently on the human researcher. This disposition, treating a result as a claim to be verified rather than a fact to be accepted, becomes more rather than less important as AI does more of the implementation work.

\textbf{Precise Task Specification.} At present, task specification should come from a human. In many experiments, leaving the choice of metric and dataset to the model worked well. However, two situations required more explicit guidance. First, when the paper proposes a library rather than reporting a performance benchmark, the main comparison table covers functional capabilities rather than running time; without an explicit directive to optimise a named function's wall-clock time, the model may pursue improvements to output quality rather than to the computational performance of the library function. Second, when a paper's comparison table includes results from multiple algorithms and the proposed method does not rank best, the prompt instruction to "outperform the reported results" is ambiguous: the model may target the highest reported value in the table rather than the paper's own proposed method.

\textbf{Novelty.} Claude Code cannot reliably check whether a proposed technique already exists in the literature. In one experiment, the model proposed using dynamic programming to solve a  problem, noting that the objective function is segment-separable and that Bellman's principle therefore applies. The plan document stated that no prior work in this literature uses DP for this problem. That claim may well be correct, but verifying it requires a systematic survey of the relevant field and judgment about which venues and search terms are authoritative. AI can assist by retrieving and summarising related work. The human researcher remains responsible for staking and defending novelty claims.

\textbf{Ethical Responsibility.} AI tools can generate code, analyze data, and draft text at a speed and scale that creates new opportunities for research misconduct, whether deliberate or inadvertent. Transparency is the first obligation. Researchers must disclose AI assistance in any submission that benefited from it. Accountability is the second. Any false claim in the work remains the author's responsibility. So does plagiarism. AI-generated text may silently reproduce passages from its training data, and an author who submits that text without verification bears full responsibility. Neither of these obligations can be delegated to the AI.

\textbf{Impact assessment.} The improvement prompt instructs Claude Code to stop immediately once it surpasses the paper's reported performance. A result that improves the metric by a fraction of a percent through parameter tuning is technically a surpass, yet it may not be scientifically interesting. Conversely, a result that reveals an unexpected structural property of the problem, for example demonstrating that an objective function is separable and therefore admits an exact polynomial-time solution where the baseline uses a metaheuristic, may be highly valuable even if the headline metric gain at small scales is modest.

\textbf{Formulating new problems.} Our pipeline is designed to improve solutions to existing, well-defined problems. Formulating a new impactful research problem is a fundamentally different and more demanding task. It requires identifying gaps or contradictions in a body of literature, recognising when a problem from one domain maps onto an unsolved problem in another, and judging which open questions are likely to be tractable. These are creative acts that depend on broad knowledge, intellectual curiosity, and intuition. AI can assist in this process, but the creative judgment about which direction is worth pursuing remains a distinctly human contribution.

\textbf{Resource provision and environment management.} In this experiment, the agent wrote code, downloaded code and datasets, and installed libraries. However, some resources could not be obtained autonomously. In one experiment, API keys had to be obtained manually and supplied to the agent. The human researcher must also ensure that sufficient computational resources such as memory, storage, and processing capacity are available before a session begins.

\textbf{Controlling Risk.} As AI coding assistants gain the ability to execute code and interact with file systems in a researcher's machine, the human researcher must actively monitor what AI does. AI agents asked for approval to execute script, access folder, installing and executing code from third-party packages.  These actions introduce risks to data and the system. Researchers must manage these risks: backing up data and code before a session begins, applying the principle of least privilege to limit what the agent can access, and monitoring its actions continuously.

\textbf{Monitoring and Directing in the research.} A complementary practice is monitoring the computational resources the model consumes during execution. On macOS, opening Activity Monitor during a session allows the researcher to verify that the process is using the expected resources.  For example, a researcher should verify that an experiment is actually using the GPU as intended. Human researchers must also make judgment calls about when to redirect the AI toward a more productive approach, when to expand from a small dataset to the full benchmark, and when to stop exploring a direction that has reached diminishing returns.

\subsection{Implications for Peer Review and Publishing}

Our experiments surfaced a recurring pattern: Claude Code can find an approach to improve performance.  In more than one case, the overlooked property was, in retrospect, straightforward to derive and should in principle have been caught during peer review. This observation creates an asymmetry: authors can use AI to strengthen their work before submission, while reviewers, under confidentiality-based editorial policies, must evaluate that work without the same assistance.

Current journal and conference policies generally prohibit reviewers from submitting manuscripts under review to public AI systems, on basis of confidentiality. This policy is prudent and should be maintained. But the asymmetry it creates is real and growing. Authors who use AI-assisted development will systematically submit stronger work; reviewers without AI assistance will be less likely to catch overlooked structural properties, suboptimal algorithmic choices.

We see two complementary mechanisms that could address this asymmetry without compromising confidentiality. First, authors can post preprints prior to formal submission, making the manuscript public and thus available to AI-assisted analysis by any reader, including potential reviewers. Second, journals could adopt an opt-in policy allowing reviewers to use AI tools for the purpose of identifying methodological issues, and the reviewer's report discloses any AI assistance. Used responsibly, this could meaningfully increase the quality of both the review process and the published work.

The growing use of AI-assisted research pipelines like ours raises questions about appropriate norms for attribution and contribution. We note that the human contributions described in Section \ref{section_human_role} remain non-trivial and are not obviously diminished by the availability of AI assistance. Major academic publishers and scholarly societies require authors to disclose AI assistance across the research process as a condition of submission; Wiley notes that ``AI can be an effective productivity tool for tasks like reference formatting, language polishing, data organization, and standard code generation." \cite{wiley2025ai}, and APA requires attribution when AI is ``used to generate ideas, content, analysis, code, or research elements" \cite{apa2025ai}. We believe this model is healthy: it creates a transparent track for AI-assisted work while leaving established venues free to set their own policies.

\section{Conclusion}

We presented a two-stage pipeline for AI-assisted improvement of published algorithm implementations and applied it to eleven experiments across diverse computational fields. The pipeline achieved improvements in every case, each within a single working day. Beyond the headline results, the experiments offer a clearer picture of how roles are distributed between the AI and the human researcher.
 
In most experiments, the AI needed no suggestion about which technical direction to pursue, and in several cases it found improvements that went beyond straightforward parameter tuning. In a few experiments, however, the human redirected the AI toward a more productive target. This shifts the practical meaning of implementing an algorithm: the bottleneck is no longer writing the code but formulating what to improve and verifying that the improvement is real, general, and novel.
 
The one contribution that proved non-negotiable was critical human judgment. Results arrived quickly and looked convincing, which invited complacency. The experiments where the human intervened, whether switching datasets, questioning whether the right baseline was targeted, or catching misidentified files, all produced better and more defensible outcomes. Critical judgment was not an occasional check; it was the core human contribution.
 
A secondary finding concerns the prompts themselves. A single short paragraph was sufficient to drive up to twenty iterations of hypothesis-driven experimentation across eleven domains, though in one experiment the AI exceeded this cap on its own initiative, prioritising the performance goal over the iteration limit. The brevity that made the prompt broadly applicable also introduced ambiguity that caused at least two experiments to proceed toward the wrong target. More precise prompts would reduce these errors but would also require more domain knowledge to write. There is a tradeoff between generality and precision that future work should explore systematically.
 
Future work will attempt to publish the improvements produced by this pipeline as standalone contributions, one per baseline paper, subject to novelty verification and full experimental validation.

\section{Acknowledgement}
The experimental work described in this paper was conducted using Claude Code by Anthropic. We explicitly declare that Claude Code wrote all experimental scripts and assisted with drafting this manuscript.  
\bibliographystyle{IEEEtran}
\bibliography{ref}

\end{document}